\documentclass[aps,twocolumn,prl,showpacs]{revtex4}
\usepackage{epsfig}

\begin{document}

\title{SU(3) prediction of a new $\Lambda$ baryon}

\author{V. GUZEY}
\affiliation{Institut f{\"u}r Theoretische Physik II, Ruhr-Universit{\"a}t Bochum,
  D-44780 Bochum, Germany}
\email{vadim.guzey@tp2.rub.de}

\preprint{RUB-TP2-19/05}

\begin{abstract}

Using the approximate flavor SU(3) symmetry of strong interactions,
we argue for the existence of a new $\Lambda$ baryon with $J^P=3/2^-$, the mass around
1850 MeV, the total width of about 130 MeV, significant branching into the
$\Sigma \pi$ and $\Sigma^{\ast} \pi$ states and
a vanishingly small coupling to the $N {\overline K}$ state.  
This confirms quark model predictions suggesting a new $\Lambda$ baryon 
in the mass range between 1775 MeV and 1880 MeV with a very small coupling to the
$N {\overline K}$ state.

\end{abstract}

\maketitle

The hypothesis of the approximate flavor SU(3) symmetry of strong interactions 
proposed by Gell-Mann and Neeman in the early 60's provided an explanation for
the observed regularities in the hadron spectrum and lead to many successful 
predictions~\cite{Eightfoldway}.
In this scheme, all known hadrons belong to singlet, octet and decuplet SU(3) representations,
 which follows from the Clebsh-Gordan series for mesons and baryons,
respectively~\cite{Kokkedee,Lichtenberg},
\begin{eqnarray}
&&{\bf 3} \otimes {\bar {\bf 3}}={\bf 1} \oplus {\bf 8} \,,  \nonumber\\
&&{\bf 3} \otimes {\bf 3} \otimes {\bf 3}=
{\bf 1} \oplus {\bf 8} \oplus {\bf 8} \oplus {\bf 10} \,.
\label{eq:clebsh_gordan}
\end{eqnarray}
All states, which do not belong to {\bf 1}, {\bf 8} or  {\bf 10} and have the baryon
number $|B| \leq 1$, are called exotic~\cite{Kokkedee,Lichtenberg}.

An assumption that the flavor SU(3) symmetry of the Hamiltonian is broken by an operator,
which transforms under SU(3) like an octet member and which preserves isospin and hyperchange,
results in the celebrated Gell-Mann-Okubo (GMO) mass formulas, which describe the mass splitting
within a given SU(3) multiplet. The GMO relations  for octets and decuplets read
\begin{eqnarray}
&&\frac{1}{2}\left( m_N+m_{\Xi} \right)=\frac{1}{4}\left(3\, m_{\Lambda}+m_{\Sigma} \right)
 \,,  \nonumber\\
&&m_{\Sigma}-m_{\Delta}=m_{\Xi}-m_{\Sigma}=m_{\Omega}-m_{\Xi}  \,.
\label{eq:gmo}
\end{eqnarray}
It is an empirical observation that the GMO mass relations work with surprisingly
high accuracy (at the level of few percent), which is much higher than the expected 20-30\%
accuracy of the SU(3) approximation.
In addition to the mass splitting, SU(3) makes rather accurate predictions for such
 properties of baryons as magnetic moments and widths of semi-leptonic weak 
decays~\cite{Lichtenberg,Garcia:1985xz}.

One can further test SU(3) by studying two-body hadronic decays. Making an assumption that
SU(3) is exact for the coupling constants
and that it is violated only by non-equal masses of hadrons,
one obtains certain relations among two-body hadronic partial decay widths of
a given multiplet, which can be confronted with the experimental data.
The most recent comprehensive analysis of all known hadrons
using the GMO mass formulas and the hypothesis of SU(3)-symmetric coupling constants was
performed by Samios, Goldberg and Meadows in 1974~\cite{Samios:1974tw}. In total, 
the authors established ten baryon multiplets and three meson nonets (a nonet is an 
octet strongly mixed a singlet) with the conclusion that {\it the detailed study of mass
relationships, decay rates, and interference phenomena shows remarkable
agreement  with that expected from the most simple unbroken SU(3) symmetry
scheme}~\cite{Samios:1974tw}. 

Recently, Guzey and Polyakov~\cite{Guzey_Polyakov}
 repeated the analysis~\cite{Samios:1974tw} for baryons
using the present knowledge of the baryon spectrum~\cite{Eidelman:2004wy}
with the conclusion that the SU(3) scheme works remarkably well.
The authors systematized virtually all known baryons with mass less than approximately
2000-2200 MeV and established twenty SU(3) baryon multiplets.
In order to have complete multiplets, a number of strange particles was predicted.
The most remarkable among them is the $\Lambda$ baryon with $J^P=3/2^{-}$, 
the mass around 1850 MeV, the total width at the level of 130 MeV and a vanishingly
 small coupling to the $N {\overline K}$ state.
The very small coupling constant to the $N {\overline K}$ state
 explains why this $\Lambda$ has not been seen in the partial wave analyses of
kaon-nucleon scattering data~\cite{Gopal:1976gs,Cameron:1977jr,Gopal:1980ur}.
Our analysis gives a model-independent confirmation of the constituent quark model prediction
that  there should exist a new $\Lambda$ baryon with the mass between 1775 MeV and 1880 MeV,
which almost decouples from the $N {\overline K}$ 
state~\cite{Isgur:1978xj,Loring:2001ky,Glozman:1997ag}.

It is important to emphasize that the existence of the new $\Lambda$ hyperon is a 
model-independent result which is a direct consequence of SU(3). This can be seen from the 
following arguments. 
Since various SU(3) multiplets have different total angular momenta, one has to 
consider representations of the SU(3)$\times$SU(2) group, where SU(3) corresponds to
flavor and SU(2) corresponds to spin. In practice, it is more convenient to study the 
representations of the larger SU(6) group, which contains SU(3)$\times$SU(2).
Taking three quarks in the fundamental SU(6) representation,
the Clebsh-Gordan series for the baryon wave function reads
\begin{equation}
{\bf 6} \otimes {\bf 6} \otimes {\bf 6}={\bf 20} \oplus {\bf 56} \oplus {\bf 70} \oplus {\bf 70} \,,
\label{eq:666}
\end{equation}
where ${\bf 20}$ is totally antisymmetric, ${\bf 56}$ is symmetric
and ${\bf 70}$ has mixed symmetry. In the following, we shall concentrate on the ${\bf 70}$
representation, which is believed to contain all negative parity 
baryons~\cite{Kokkedee,Lichtenberg,Samios:1974tw,Guzey_Polyakov}.
A standard textbook  analysis~\cite{Kokkedee,Lichtenberg} shows
that ${\bf 70}$ has the following SU(3)$\times$SU(2) content
\begin{equation}
{\bf 70}=\left({\bf 1},\ \frac{1}{2} \right) \oplus \left({\bf 8},\ \frac{1}{2} \right)
\oplus \left({\bf 8},\ \frac{3}{2} \right) \oplus \left({\bf 10},\ \frac{1}{2} \right) 
\,, \
\label{eq:su6tosu3}
\end{equation} 
where the second quantity in the parenthesis denotes the total spin $S$ of the multiplet.
The final baryon wave function is obtained by coupling the total orbital 
moment of three quarks $L$ to their spin $S$. For the negative parity baryons, one uses
$L=1$ and obtains the following list of possible negative parity SU(3) multiplets
\begin{eqnarray}
({\bf 70},L=1)&=&\left({\bf 1},\frac{1}{2} \right) \oplus \left({\bf 1}, \frac{3}{2} \right) \oplus 2 \left({\bf 8}, \frac{1}{2} \right) 
\oplus2  \left({\bf 8},\frac{3}{2} \right) \nonumber\\
& \oplus& \left({\bf 8}, \frac{5}{2} \right) \oplus \left({\bf 10},\frac{1}{2} \right)\oplus  \left({\bf 10}, \frac{3}{2} \right)\,,  
\label{eq:counting}
\end{eqnarray} 
where now the second quantity in the parenthesis denotes the total angular momentum $J$
of the multiplet. Equation~(\ref{eq:counting}) unambiguously states that one must 
observe at least seven negative-parity $\Lambda$
baryons (the SU(3) singlets must be $\Lambda$'s).  Since the 
Review of Particle Physics (RPP) contains only six negative
parity $\Lambda$ in the required mass range~\cite{Eidelman:2004wy}, one  
$\Lambda$ with $J^P=3/2^-$ is clearly missing. 
Since all other $\Lambda$ baryons, which are required by SU(3) for positive and negative
parity multiplets,  are very well established and have a very high status in the RPP
 (three and four stars), the missing $\Lambda$ stands out very dramatically as a very 
important missing piece of the whole SU(3) picture of baryon multiplets.

Let us now examine the $J^P=3/2^-$ octet with the missing $\Lambda$ in detail.
The octet opens with the well-established three-star $N(1700)$. 
In addition, in the appropriate mass
range, one can offer a candidate for the 
$\Sigma$ member of the considered octet -- the three-star
$\Sigma(1940)$. Since the octet in question lacks two states, one
 cannot use the GMO mass formula~(\ref{eq:gmo}) to estimate the mass of the missing $\Lambda$. 
Instead, we notice that, on average, 
the mass difference between the $N$ and $\Lambda$ states is approximately 150 MeV.
 Therefore, we {\it assume} that the mass of the missing $\Lambda$ hyperon
is around 1850 MeV. Note that the mass of $N(1700)$, which we use
as a reference point, is itself
known with a large uncertainty: $m_{N(1700)}=1650-1750$ MeV according to the 
RPP estimate~\cite{Eidelman:2004wy}. Therefore, the uncertainty in the predicted 
mass of $\Lambda(1850)$ is approximately 50 MeV.
 
The mass of the missing $\Xi$ member of the octet is estimated using the GMO mass formula.
 Using $m_N=1700$ MeV, $m_{\Sigma}=1940$ MeV and $m_{\Lambda}=1850$ MeV, 
 we obtain $m_{\Xi}=2045$ MeV.

In order to make further predictions about the properties of the
octet and the predicted $\Lambda$, 
we consider two-body hadronic decays using the assumption 
of SU(3)-symmetric coupling constants~\cite{Samios:1974tw}. In this limit, 
the $g_{B_1 B_2 P}$ coupling constants of $B_1 \to B_2 +P$ decays
 ($B_2$ belongs to the ground-state octet or decuplet; $P$ belongs to the octet
 of pseudoscalar mesons) 
can be parameterized in terms of three free parameters: $A_s$ and $A_a$ for 
${\bf 8} \to {\bf 8}+{\bf 8}$ decays and $A_8^{\prime}$ for
${\bf 8} \to {\bf 10}+{\bf 8}$ decays,
\begin{eqnarray}
g_{B_1 B_2 P }&=&A_s \left(
\begin{array}{cc}
8 & 8 \\
Y_2 I_2 & Y_P I_P
\end{array}\right|\left.\begin{array}{c}
          8_S\\Y_1 I_1
          \end{array}\right) \nonumber\\
&+&A_a \left(
\begin{array}{cc}
8 & 8 \\
Y_2 I_2 & Y_P I_P
\end{array}\right|\left.\begin{array}{c}
          8_A\\Y_1 I_1
          \end{array}\right)  \,,\nonumber\\
g_{B_1 B_2 P }&=&A_8^{\prime}\left(
\begin{array}{cc}
10 & 8 \\
Y_2 I_2 & Y_P I_P
\end{array}\right|\left.\begin{array}{c}
          8\\Y_1 I_1
          \end{array}\right) \,.
\label{eq:888}
\end{eqnarray}
Note that there are two coupling constants $A_s$ and $A_a$ because
the tensor product ${\bf 8} \otimes {\bf 8}$ contains one  
symmetric and one antisymmetric octet representation,
${\bf 8}_S$ and ${\bf 8}_A$~\cite{Kokkedee,Lichtenberg}.
 In Eq.~(\ref{eq:888}), $Y$ and $I$ denote the hypercharge and isospin of the
involved hadrons; the factors multiplying the coupling
constants are the so-called  SU(3) isoscalar factors~\cite{deSwart:1963gc}.
For practical applications, it is convenient to use an 
alternative pair of parameters $A_8$ and $\alpha$~\cite{Samios:1974tw}
\begin{equation}
A_8=\frac{\sqrt{15}}{10} A_s+\frac{\sqrt{3}}{6} A_a \,, \quad \quad \alpha=\frac{\sqrt{3}}{6} \frac{A_a}{A_8} \,.
\end{equation}
The coupling constants of Eq.~(\ref{eq:888}) for selected relevant decay modes are 
summarized in Table~\ref{table:octets}.
\begin{table}[h]
\begin{center}
\begin{tabular}{|c c c|}
\hline
Decay mode & $g_{B1 B_2 P}$  & $g_{B1 B_2 P}$\\
\hline
$N \to N \pi$ & $\sqrt{3}\, A_8$  & \\
$ \to N \eta$ & $[(4\, \alpha-1)/\sqrt{3}]\, A_8$  &  \\
$ \to \Sigma K$ & $\sqrt{3}\, (2 \,\alpha-1)\, A_8$  & \\
$ \to \Lambda K$ & $-[(2 \alpha+1)/\sqrt{3}]\, A_8$  & \\
$ \to \Delta \pi$ & & $-2/\sqrt{5} \, A_8^{\prime}$ \\
\hline
$\Lambda \to N {\overline K}$ & $\sqrt{2/3}\,(2\, \alpha+1) \, A_8$  & \\
$\to \Sigma \pi$ & $2\,(\alpha-1) \, A_8$  &  \\
$\to \Lambda \eta$ & $2/\sqrt{3}\,(\alpha-1) \, A_8$ & \\
$ \to \Sigma^{\ast} \pi$ &  & $-\sqrt{15}/5 \, A_8^{\prime}$  \\
\hline
$\Sigma \to \Sigma \pi$ & $2 \sqrt{2} \, \alpha \,A_8$  & \\
$\to \Lambda \pi$ &  $-2/\sqrt{3}\,(\alpha-1) \, A_8$  &  \\
$\to N {\overline K}$ & $\sqrt{2}\,(2 \,\alpha-1) \, A_8$  &  \\
$\to \Delta {\overline K}$ &  & $2 \sqrt{30}/15 \, A_8^{\prime}$   \\
$\to \Sigma^{\ast} \pi$ &  & $-\sqrt{30}/15 \, A_8^{\prime}$   \\
\hline
$\Xi \to \Xi \pi$ & $\sqrt{3}\,(2\, \alpha-1) \, A_8$  &  \\
$\to \Lambda {\overline K}$ & $[(4 \,\alpha-1)/\sqrt{3}]\, A_8$  & \\
$\to \Sigma {\overline K}$ & $\sqrt{3}\, A_8$  &  \\
$ \to \Xi^{\ast} \pi$ &  & $-\sqrt{5}/5 \, A_8^{\prime}$ \\
$ \to \Sigma^{\ast} {\overline K}$ & &  $\sqrt{5}/5 \, A_8^{\prime}$ \\
\hline
\end{tabular}
\caption{The selected SU(3) universal coupling constants for ${\bf 8} \to {\bf 8}+{\bf 8}$
and  ${\bf 8} \to {\bf 10}+{\bf 8}$ decays.}
\label{table:octets}
\end{center}
\end{table}

It is important to note that while the $A_8$ and  $A_8^{\prime}$ coupling constants 
are totally free parameters, SU(6) makes predictions for the parameter  $\alpha$
because it is related to the so-called $F/D$ ratio. For the considered octet,
$\alpha^{{\rm SU(6)}}=-1/2$~\cite{Samios:1974tw}.

The SU(3) prediction for the partial decay width has the form~\cite{Samios:1974tw}
\begin{equation}
\Gamma\left(B_1\to B_2+P \right)=\left|g_{B_1 B_2 P} \right|^2 
\left(\frac{k}{M}\right)^{2l} \left(\frac{k}{M_1}\right) M \,,
\label{eq:width}
\end{equation}
where $k$ is the center-of-mass momentum of the final particles;
$M_1$ is the mass of $B_1$; $M=1000$ MeV is the dimensional parameter;
$l$ is the relative orbital momentum of the outgoing $B_2 \, P$ system. 
The orbital momentum $l$ is found by requiring the conservation of parity and
the total orbital momentum in the decay. Thus, $l=2$ for the ${\bf 8} \to
 {\bf 8}+{\bf 8}$ decays of the considered $J^P=3/2^-$ octet and
$l=0,2$ for the ${\bf 8} \to {\bf 10}+{\bf 8}$ decays. 
Note that with the definition~(\ref{eq:width}), the $g_{B_1 B_2 P}$ coupling
constants are dimensionless.

Using Eq.~(\ref{eq:width}), we perform the $\chi^2$ fit~\cite{MINUIT} to the
experimentally measured partial decay widths of $N(1700)$ and $\Sigma(1940)$.
The results are summarized in Table~\ref{table:m11}, where the observables used in
the fit are underlined and SU(3) predictions are listed in the last column.
Note that, by definition, the square root of the product of two decay widths can
be either positive or negative, depending of the relative phase between the
corresponding amplitudes, see Table~\ref{table:octets}. This serves 
as a stringent test of SU(3) predictions.
 One sees
that the central values of the fitted observables, which are known with very large 
experimental uncertainties, are reproduced well (except for $\Gamma_{\Sigma(1940) \to
\Delta \overline{K}}$).
\begin{table}[h]
\begin{center}
\begin{tabular}{|c c c c |}
\hline
 Mass and  & \ Observables \ & Exper. (MeV) & SU(3) (MeV) \\
 width (MeV) &  &  &  \\
\hline
$N(1700)$             & $\Gamma_{N \pi}$  &   \underline{$10.0 \pm 7.1$} &  7.9 \\
$\Gamma=100 \pm 50$   &  $\Gamma_{N \eta}$  & $0 \pm 1$ & 2.0 \\
                     &  $\Gamma_{\Delta \pi}$, $l=2$ & \underline{$14.4 \pm 17.0$} & 17.9 \\
& & & \\
$\Lambda(1850)$ &    $\Gamma_{N {\overline K}}$ & & 0.2 \\
 & $\Gamma_{\Sigma \pi}$ & & $17.8$ \\
& $\Gamma_{\Lambda \eta}$ & & 1.2 \\
& $\Gamma_{\Sigma^{\ast} \pi}$ & & $12.8$ \\
& & & \\
$\Sigma(1940)$ &  $|\sqrt{\Gamma_{N {\overline K}} \Gamma_{\Sigma \pi}}|$ & 
\underline{$24.0 \pm 13.6$} & $18.4$ \\
$\Gamma=300 \pm 80$ & $\sqrt{\Gamma_{N {\overline K}} \Gamma_{\Lambda \pi}}$ & 
\underline{$-18.0 \pm 10.2$} & $-22.2$ \\
&                    $\Gamma_{\Delta \overline{K}}$, $l=2$ & \underline{$47.2 \pm 42.0$} & 10.6 \\
& $\Gamma_{\Sigma \pi}$ & & 9.1 \\
& $\Gamma_{\Lambda \pi}$ & & 13.1 \\
&  $\Gamma_{N \overline{K}}$ & & 37.4 \\
&  $\Gamma_{\Sigma^{\ast}\pi}$ & & 5.4 \\
& & & \\
$\Xi(2045)$ & $\Gamma_{\Xi \pi}$ & & 39.5  \\
& $\Gamma_{\Lambda \overline{K}}$  & & 12.3  \\
& $\Gamma_{\Sigma \overline{K}}$  & & 4.8  \\
& $\Gamma_{\Xi^{\ast} \pi}$  & & 6.7  \\
& $\Gamma_{\Sigma^{\ast} \overline{K}}$  & & 2.6  \\
\hline
\end{tabular}
\caption{Results of the $\chi^2$ fit to two-body hadronic decays of the considered octet.}
\label{table:m11}
\end{center}
\end{table}

The input for our $\chi^2$ requires an explanation.
Since for $N(1700)$ the total width and the important branching into the $N \pi$ final state
are known with large ambiguity, we use the RPP estimates~\cite{Eidelman:2004wy}.
For  $\Sigma(1940)$, we use the results of Gopal {\it at al.}~\cite{Gopal:1976gs}.
According to the analysis~\cite{Gopal:1976gs}, both
$\sqrt{\Gamma_{N {\overline K}}\Gamma_{\Lambda \pi}}$ and  
$\sqrt{\Gamma_{N {\overline K}}\Gamma_{\Sigma \pi}}$ of $\Sigma(1940)$ are negative.
This contradicts SU(3): expecting that $\alpha$ 
is close to its SU(6) prediction $\alpha=-1/2$, we notice that SU(3) requires that the
signs of $\sqrt{\Gamma_{N {\overline K}}\Gamma_{\Lambda \pi}}$ and 
$\sqrt{\Gamma_{N {\overline K}}\Gamma_{\Sigma \pi}}$ should be opposite,
see Table~\ref{table:octets}.
SU(3) also requires the opposite signs, if $\Sigma(1940)$ belongs to a decuplet.
Therefore, we reverse the sign of $\sqrt{\Gamma_{N {\overline K}}\Gamma_{\Sigma \pi}}$.
This is consistent with the analysis~\cite{Martin:1977md}, which 
reports the positive value for $\sqrt{\Gamma_{N {\overline K}}\Gamma_{\Sigma \pi}}$,
which is somewhat larger (no errors are given) than the value from the 
analysis~\cite{Gopal:1976gs}.

The $\chi^2$ fit to the five underlined observables in 
Table~\ref{table:m11} gives
\begin{eqnarray}
A_8  & =&   8.3 \pm 3.5 \,, \ \alpha=-0.70 \pm 0.54 \,, \ \chi^2/{\rm d.o.f.}=0.42/1 \,, 
\nonumber\\
A_8^{\prime}& =& 67.2 \pm 31.0 \,, \ \chi^2/{\rm d.o.f.}=0.8/1 \,. 
\label{eq:m11}
\end{eqnarray}
The central value of $\alpha$ compares well to its SU(6) prediction.
However, since we used in the fit only two observables,
 which depend on $\alpha$, the error on the obtained value of 
$\alpha$ is large.

Note that in order to convert the experimentally measured 
$\sqrt{Br(N {\overline K}) Br(\Delta \overline{K})}$ 
of $\Sigma(1940)$ into the corresponding $\Gamma_{\Delta \overline{K}}$ 
 used in the fit, we used the SU(3) prediction 
$\Gamma_{\Sigma(1940)\to N {\overline K}}=37.4$ MeV. 
Also, we chose to fit only the $D$-wave [$l=2$ in Eq.~(\ref{eq:width})] ${\bf 8} \to {\bf 10}+{\bf 8}$ decays
because the $S$-wave ($l=0$) $N(1700) \to \Delta \pi$ branching ratio is rather uncertain and
is smaller than the corresponding $D$-wave branching~\cite{Eidelman:2004wy}.

An examination of SU(3) predictions in Table~\ref{table:m11} shows that the sum of
predicted two-body decay widths significantly underestimates the known total
widths of $N(1700)$ and $\Sigma(1940)$: $\Gamma_{N(1700)}^{{\rm SU(3)}}=28$ MeV 
vs.~$\Gamma_{N(1700)}=100 \pm 50$ MeV and
$\Gamma_{\Sigma(1940)}^{{\rm SU(3)}}=77$ MeV 
vs.~$\Gamma_{\Sigma(1940)}=300 \pm 80$ MeV. In both cases, the central value of the total width
is underestimated by the factor of $3.5-4$. Therefore,
in order to obtain a realistic estimate for the total width of the 
predicted $\Lambda(1850)$,
we multiply the sum of the SU(3) predictions for the two-body hadronic decays
 by the factor four
\begin{equation}
\Gamma_{\Lambda(1850)}=4 \times 32 \ {\rm MeV} = 128 \ {\rm MeV} \,.
\label{eq:mass_lambda}
\end{equation}
We can estimate the total
width of the predicted $\Xi(2045)$ in a similar way
\begin{equation}
\Gamma_{\Xi(2045)}=4 \times 66 \ {\rm MeV} = 264 \ {\rm MeV} \,.
\label{eq:mass_xi}
\end{equation}

Using the octet dominance hypothesis of Gell-Mann, relations among 
total widths of baryons of the same SU(3) multiplet, which look identical to the GMO mass
splitting formulas, were derived by Weldon in 1978~\cite{Weldon:1977wf}.
The Weldon's relation for octets reads
\begin{equation}
\frac{1}{2}\left(\Gamma_N+\Gamma_{\Xi} \right)=
\frac{1}{4}\left(3\, \Gamma_{\Lambda}+\Gamma_{\Sigma} \right) \,.
\label{eq:weldon}
\end{equation}
Using the total widths from Table~\ref{table:m11} and from
Eqs.~(\ref{eq:mass_lambda}) and (\ref{eq:mass_xi}), we observe that
the Weldon's relation for the considered octet is satisfied with high accuracy
\begin{eqnarray}
\frac{1}{2}\left(\Gamma_N+\Gamma_{\Xi} \right)& = & 182 \ {\rm MeV} \,, \nonumber\\
 \frac{1}{4}\left(3\, \Gamma_{\Lambda}+\Gamma_{\Sigma} \right) & = & 171 
\ {\rm MeV} \,.
\label{eq:agreement}
\end{eqnarray}
While the nice agreement seen in Eq.~(\ref{eq:agreement}) should not be 
taken literally because of large uncertainties in the measured total widths
of $N(1700)$ and $\Sigma(1940)$, it still illustrates that our method of
estimating the total widths of $\Lambda(1850)$ and $\Xi(2045)$ has
a certain merit.

A remarkable property of $\Lambda(1850)$ is its vanishingly small coupling
to the $N \overline{K}$ final state, see Table~\ref{table:m11}. This is a consequence
of the fact that $\alpha=-0.70 \pm 0.54$, which strongly suppresses the
$g_{\Lambda(1850) \to N \overline{K}}$ coupling constant,
$g_{\Lambda \to N \overline{K}}=\sqrt{2/3} (2 \alpha +1) A_8$, 
see Table~\ref{table:octets}. In the SU(6) limit, $\alpha=-1/2$, which leads to 
$g_{\Lambda \to N \overline{K}}=0$.
Therefore, our prediction that $\Lambda(1850)$ very weakly couples to the
$N \overline{K}$ final state is rather model-independent.

As follows from Table~\ref{table:m11}, SU(3) predicts that the $\Lambda(1850)$ 
has significant branching ratios into the
$\Sigma \pi$ and $\Sigma^{\ast} \pi$ final states. This suggests that
one should experimentally search for the $\Lambda(1850)$
in production reactions using the $\Sigma \pi$ and $\Sigma^{\ast} \pi$ invariant
mass spectrum.

The existence of a new $\Lambda$ hyperon with $J^P=3/2^-$ was predicted in
different constituent quark models. 
In 1978, Isgur and Karl predicted that the new $\Lambda$
has the mass 1880 MeV and very small coupling to the $N \overline{K}$ state. The latter
fact is a consequence of SU(6) selection rules and explains why this state was not
observed in the $N \overline{K}$ partial wave 
analyses~\cite{Gopal:1976gs,Cameron:1977jr,Gopal:1980ur}.
In a subsequent analysis, Isgur and Koniuk explicitly calculated
the partial decay widths of the  $\Lambda(1880)$ and found that 
$\Gamma_{N \overline{K}}$ is small, while $\Gamma_{\Sigma \pi}$ and
$\Gamma_{\Sigma^{\ast} \pi}$ are dominant~\cite{Koniuk:1979vy}.

More recent calculations within the  constituent quark model framework 
also predict the existence of a new $\Lambda$ with $J^P=3/2^-$, but with somewhat different
masses: the analysis of L\"oring, Metsch and Petry~\cite{Loring:2001ky} (model A)   
gives 1775 MeV; the analysis of  Glozman, Plessas, Varga and Wagenbrunn
gives $\approx 1780$ MeV. Note also that the analysis~\cite{Loring:2001ky}
predicts that the $\Lambda$ very weakly couples to the  $N \overline{K}$ state.

We would like to emphasize that while many results concerning the new $\Lambda$ were
previously derived in specific constituent quark models with various
assumptions about the quark dynamics, we demonstrate that they
are actually model-independent and follow directly from
flavor SU(3) symmetry.

In conclusion, the existence of a new $\Lambda$ hyperon with $J^P=3/2^-$ is
required by the general principle of the flavor SU(3) symmetry of strong 
interactions. Our SU(3) analysis predicts that its mass is $\approx 1850$ MeV
 and the total width is $\approx 130$ MeV. We predict that 
$\Lambda(1850)$ has very small coupling to the $N \overline{K}$ state and 
large branching into the $\Sigma \pi$ and $\Sigma^{\ast} \pi$ final states.   
Therefore, $\Lambda(1850)$ can be searched for in production reactions by
studying the $\Sigma \pi$ and $\Sigma^{\ast} \pi$ invariant mass spectra. 
The fact that the total width of
$\Lambda(1850)$ is not larger than $\approx 130$ MeV makes the experimental search feasible.

The author thanks M.V.~Polyakov for useful discussions and collaboration
and M.~Strikman for useful comments on the manuscript.  This work was 
supported by the S.~Kovalevskaya Program of the Alexander von Humboldt Foundation.

\end{document}